\documentclass[12pt,aaspp4,tighten,flushrt]{article}
\usepackage{euscript}

\def\ni{\noindent}
\def\.{\mathaccent 95}
\def\beq{\begin{equation}}
\def\ee{\end{equation}}

\def\be{\beta}

\def\ep{\epsilon}

\def\la{\lambda}

\def\frac#1#2{{\textstyle{{#1}\over {#2}}}}
\def\ni{\noindent}
\def\lsim{\mathrel{\rlap{\lower4pt\hbox{\hskip1pt$\sim$}}
    \raise1pt\hbox{$<$}}}
\def\gsim{\mathrel{\rlap{\lower4pt\hbox{\hskip1pt$\sim$}}
    \raise1pt\hbox{$>$}}}
\def\sqr#1#2{{\vcenter{\vbox{\hrule height.#2pt
         \hbox{\vrule width.#2pt height#1pt \kern#1pt
         \vrule width.#2pt}
         \hrule height.#2pt}}}}

\newbox\grsign \setbox\grsign=\hbox{$>$} \newdimen\grdimen \grdimen=\ht\grsign
\newbox\simlessbox \newbox\simgreatbox
\setbox\simgreatbox=\hbox{\raise.5ex\hbox{$>$}\llap
     {\lower.5ex\hbox{$\sim$}}}\ht1=\grdimen\dp1=0pt
\setbox\simlessbox=\hbox{\raise.5ex\hbox{$<$}\llap
     {\lower.5ex\hbox{$\sim$}}}\ht2=\grdimen\dp2=0pt

\def\x      {{\hbox{X-ray}}}

\def\etal   {{\it et~al.}}

\def\doublespace {\smallskipamount=6pt plus2pt minus2pt
                  \medskipamount=12pt plus4pt minus4pt
                  \bigskipamount=24pt plus8pt minus8pt
                  \normalbaselineskip=24pt plus0pt minus0pt
                  \normallineskip=2pt
                  \normallineskiplimit=0pt
                  \jot=6pt
                  {\def\smallskip {\vskip\smallskipamount}}
                  {\def\medskip   {\vskip\medskipamount}}
                  {\def\bigskip   {\vskip\bigskipamount}}
                  {\setbox\strutbox=\hbox{\vrule 
                    height17.0pt depth7.0pt width 0pt}}
                  \parskip 12.0pt
                  \normalbaselines}

\font\gkvec=cmmib10                         
\def\bomega{\hbox{{\gkvec\char33}}}                  

\def\lb{\langle}
\def\rb{\rangle}

\def\bfJ{\bf J}
\def\bfA{\bf A}
\def\bfa{\bf a}
\def\bbE{\overline {\bf E}}
\def\bbP{\overline {\bf P}}
\def\bw{\overline {\omega}}
\def\bv{\overline V}
\def\bB{\overline B}
\def\ts{\times}
\def\lb{\langle}
\def\rb{\rangle}
\def\curl{\nabla \ts}

\def\bbV{\overline {\bf V}}
\def\bfv{{\bf v}}

\def\bfV{{\bf V}}
\def\bfj{{\bf j}}

\def\bfe{{\bf e}}
\def\bfE{{\bf E}}
\def\bfw{{\bomega}}

\def\bfb{{\bf b}}

\def\bfB{{\bf B}}

\def\bbB{\overline{\bf B}}
\def\bbb{\overline{\bf B}}
\def\bbJ{\overline{\bf J}} 
\def\bbA{\overline{\bf A}} 
\def\bbE{\overline{\bf E}} 
\def\nb{\nabla}
\def\curl{\nb\ts}
\def\div{\nb\cdot}
\def\b0{b^{(0)}}
\def\v0{v^{(0)}}
\def\w0{\omega^{(0)}}
\def\bb0{\bfb^{(0)}}
\def\bv0{\bfv^{(0)}}
\def\bw0{\bfw^{(0)}}
\def\bj0{\bfj^{(0)}}

\def\ni{\noindent}
\begin{document}

\def\be{\begin{equation}}
\def\ee{\end{equation}}
\def\lab#1{\label{#1}}
\def\lrp#1{\left(#1\right)}
\def\BV{{\bf V}}
\def\etal{{\it et al.}\ }
\def\OV{\overline{\bf V}}
\def\E{{\bf E}}
\def\x{{\bf x}}

\def\pref#1{(\ref{#1})}
\def\beq{\begin{eqnarray}}
\def\eeq{\end{eqnarray}}
\def\nn{\nonumber}
\def\nt{\nabla\times}
\def\OE{\overline{\bf E}}
\def\A{{\bf A}}
\def\lra#1{\left\langle #1\right\rangle}
\def\bv{\bf v}
\def\OB{\overline{\bf B}}
\def\ove{\overline{E}}
\def\cnt{\cdot\nabla\times}
\def\b{{\bf b}}
\def\ob{\overline{B}}
\def\ao{\alpha\hbox{-}\Omega}
\def\B{{\bf B}}

\centerline {\bf HOW ASTROPHYSICAL MEAN FIELD DYNAMOS CAN}
\centerline {\bf CIRCUMVENT EXISTING QUENCHING CONSTRAINTS}

\medskip
\centerline{Eric G. Blackman, Department of Physics \& Astronomy} 
\centerline{University of Rochester, Rochester NY 14627} 
\centerline {and}
\centerline {George B. Field, Harvard-Smithsonian 
Center for Astrophysics (CFA)}
\centerline {60 Garden St., Cambridge MA, 02139, USA}

\bigskip 
\medskip
\centerline{\bf Abstract}
\medbreak

Mean field dynamo  theory is a leading candidate 
to explain the large scale magnetic flux 
in galaxies and stars.  However, 
controversy arises over the extent of premature quenching 
by the backreaction of the growing field. 
We distinguish between rapid dynamo action, which is required by
astrophysical systems, and resistively limited
dynamo action.  We show how the flow of magnetic helicity is important for rapid dynamo action. Existing numerical and 
analytic work suggesting that mean field dynamos are prematurely quenched and 
resistively limited include approximations or boundary 
conditions which suppress the magnetic helicity flow from the outset.
Thus they do not unambiguously reveal whether flux generating 
astrophysical dynamos are 
dynamically suppressed when the required helicity flow is allowed.
An outflow of helicity also implies an outflow of magnetic energy and so
active coronae and/or winds are a prediction of mean field dynamos.  
Open boundaries alone may not be sufficient and the additional physics of 
buoyancy and winds may be required. 
Some possible simulation approaches, even with periodic boxes,
to test the principles are discussed.  
Some imitations of the ``Zeldovich relation'' are
also addressed.


\vfill
\eject

\ni {\bf I. INTRODUCTION}
\bigskip


A leading candidate to explain
the origin of large scale magnetic flux growth in stars and galaxies 
has been the mean field dynamo (MFD) theory 
[1,2,3,4,5,6]. The theory appeals to a combination of helical
turbulence (leading to the $\alpha$ effect), differential rotation (the $\Omega$
effect),  and turbulent diffusion to exponentiate an initial seed mean
magnetic field. 
Ref [7] developed a
formalism for describing the concept [8] that helical turbulence  
can twist toroidal ($\phi$)  fields 
into the poloidal ($r,z$) direction, where they can be
acted upon by differential rotation to regenerate a powerful large scale
toroidal magnetic field.  Their formalism involved breaking the total
magnetic field into 
a mean 
component $\OB$ and a fluctuating  component $\b$, and similarly for the
velocity field $\BV$.  The mean can be a spatial mean or an ensemble
average.  For comparison to observations of a single astrophysical 
system, the ensemble average
is approximately equal to the spatial average when there is a 
scale separation between the mean scale and the fluctuating scale.
In reality, the scale separation is often weaker than the  
dynamo theorist desires. Nevertheless, we  proceed to consider 
spatial averages.

Ref [7] showed that $\OB$ satisfies the induction 
equation
\be
{\partial\OB\over \partial t} = -c\curl\bbE,
\lab{2.4a} 
\ee
where
\be
\bbE=-\lrp{\OV\times \OB}/c  - \lb\bfv\ts\bfb\rb/c+
\lambda\nt\OB,
\lab{2.4aa} 
\ee
where the first term describes the effect of differential rotation
(``$\Omega$-effect"),
\be
\lb\bfv\ts\bfb\rb_i = \alpha_{ij}{\bB}_j-\beta_{ijk}\partial_j{\bB}_k
\lab{2.5a}
\ee
is the ``turbulent emf," and $\lambda=\eta c^2/4\pi$ is the magnetic
diffusivity defined with the resistivity $\eta$.
Here $\alpha_{ij}$ contains Parker's twisting (``$\alpha$ effect")
and $\beta_{ijk} (\gg \lambda)$ contains the turbulent diffusivity. Ref [7]
calculated $\OE$ to first order in $\ob$ for isotropic
$\alpha_{ij}$ and $\beta_{ijk}$ and hence the dynamo
coefficients $\alpha$ and 
$\beta$ to zeroth order in $\ob$ from the statistics of the turbulence.
They ignore the Navier-Stokes equation. 
The back-reaction on the dynamo coefficients to first
order in $\overline {\bf B}$ and $\overline {\bf V}$ was calculated
in Ref [9], and Ref [10] calculates $\alpha$ 
to all orders in $\ob$ when mean field gradients are small)
See also Refs [11, 12] and the issues raised in [13].
Refs [11,12] obtain a catastrophically quenched $\alpha$ 
but by a different argument than we use later.

When (\ref{2.4aa}) is substituted into (\ref{2.4a}), 
we have the mean-field dynamo equation:
\be
{\partial\OB\over \partial t} = \nt \lrp{\OV\times \OB} 
+\curl(\alpha \bbB)-\curl(\beta +\la)\curl \OB. \lab{6}
\ee
In the approximation that $\overline {\bf V}$, 
$\alpha$ and $\beta$ are independent of $\ob$, (\ref{6}) 
is a linear equation for $\OB$ which can be solved as an eigenvalue
problem for the growing modes in the Sun and other bodies.
Actually, a rapid growth of the fluctuating field necessarily accompanies the 
mean-field dynamo.  Its impact upon the growth of the mean
field, and the impact of the mean field itself on its own growth are
controversial.  

The controversy results because  
Lorentz forces from the growing magnetic field react back on and
complicate the turbulent  motions driving the field growth  
[14,15,16,17,18,19]. It is tricky to
disentangle the back reaction of the mean field from that of the 
fluctuating field. Analytic studies and numerical
simulations seem to disagree as to the extent to which the dynamo
coefficients are suppressed by the back reaction of the mean field. 
But, as we will address,  
the disagreements may be because different problems are being solved 
and the results inappropriately compared. 

It is important to distinguish between rapid MFD
action, resistively limited MFD action, and no MFD action.
Rapid dynamo action describes dynamo action which proceeds at rates
not greatly suppressed from the kinematic values.
Resistively limited dynamo action is MFD growth that 
proceeds at rates strongly dependent on the magnetic Reynolds number.
No MFD action implies no mean growth at all.
For galaxies, rapid MFD action is necessary if the observed large
scale fields are to be produced and sustained by the MFD.
There the $\alpha$ effect and the $\Omega$ effect operate in the same
volume. For the Sun, rapid dynamo action is also required but 
the $\alpha$ effect operates in the convection zone, whilst
the $\Omega$ shear effect operates in the overshoot layer beneath
the convection zone [20]. 

In section II we discuss the role of magnetic helicity
in dynamo theory. We show 
that open boundaries, allowing magnetic helicity
to escape, play an important role for rapid astrophysical dynamo action. 
This leads to an ambiguity in interpreting of quenching simulations.
In section III we predict that astrophysical rotators 
with dynamos harbor steady coronae. 
In section IV we suggest that open boundary conditions
may not be enough for a dynamo and that the magnetic helicity flow 
may require  buoyancy or winds.  We also address pitfalls
of the Zeldovich relation. We conclude in section V.



\bigskip
\ni {\bf II. ROLE OF MAGNETIC HELICITY CONSERVATION}
\bigskip

Although the MFD theory 
predates detailed studies of MHD turbulence, the MFD may be looked
upon in hindsight as a framework for 
studying the inverse cascade of magnetic helicity.
Whether this inverse cascade is primarily local (proceeding by
interactions of eddies/waves of nearby wavenumbers) or non-local
(proceeding with a direct conversion of power from 
large to small wavenumbers) is important to understand.
The simple MFD  is most consistent with the latter.

From the numerical
solution of approximate equations describing the spectra of energy and
helicity in MHD turbulence, Ref [21] showed 
that the $\alpha$ effect conserves magnetic
helicity $(=\int({\bf A}\cdot{\bf B})d^3x)$
by pumping a positive (negative) amount to scales $>L$ (the outer
scale of the turbulence) while pumping a negative (positive) amount to
scales $\ll L$. Magnetic 
energy at the large scale was identified with the $\OB$ of [7].  Thus,
dynamo action leading to an ever larger $\ob$, hence the creation of ever
more large scale helicity, can proceed as long as
small scale helicity of opposite sign can be removed or dissipated.
Recent simulations [22] confirm this inverse cascade
and the role of $H_M$ conservation.  But the rate of small scale $H_M$ removal 
determines the rate of MFD action. Presently, simulations have invoked
boundary conditions for which the growth of large scale field is 
resisitvely limited.  We now discuss what this means.


\bigskip
{\bf A. Constraint equation for the turbulent EMF}
\bigskip


To explore the role of boundary conditions
in constraining the value of the $\alpha$ dynamo parameter, 
we take Ohm's law
\be
\bfE= {-c^{-1}\bfV \times \bfB +\eta {\bf J}} 
\label{ohm}
\ee
and average the dot product with $\bfB$ to find [23]
\be
\lb{\bfE \cdot \bfB}\rb 
=\bbE\cdot \bbB+\lb{\bfe\cdot \bfb}\rb 
= -c^{-1}\lb{\bfv\times \bfb}\rb\cdot \bbB+\eta \bbJ\cdot \bbB+
\lb{\bfe\cdot \bfb}\rb 
\lab{mon21}
\ee
where ${{\bf J}}$ is the current
density.

A second expression for $\lb{\bf E}\cdot {\bf B}\rb$ also follows from 
Ohm's law without first splitting into mean and fluctuating components,
that is 
\be
\lb{\bfE\cdot \bfB}\rb 
=\eta \lra{\bfJ\cdot \bfB} = \eta \bbJ\cdot \bbB+\eta
\lb{\bfj \cdot \bfb}\rb =\eta\bbJ\cdot \bbB 
+c^{-1} \lambda \lra{\bfb\cdot\nabla\times \bfb}.
\lab{mon22}
\ee
Using  
(\ref{mon22}) and  (\ref{mon21}), we have 
\be
-c^{-1} \lb{\bfv\times \bfb}\rb \cdot \bbB 
 = c^{-1} \la \lra{\bfb\cdot
\nb\times \bfb } -\lb{\bf e}\cdot {\bf b}\rb, 
\lab{mon23}
\ee
which can be used to constrain  $\lb\bfv\ts\bfb\rb$ in the mean field
theory.

\bigskip
 {\bf B. Necessity of magnetic helicity escape and open boundaries}
\bigskip

Now consider ${\bf E}$ in terms of the vector 
and scalar potentials ${\bf A}$ and $\Phi$:
\begin{equation}
\bfE=-\nabla \Phi -(1/c)\partial_t {\bf A}.
\label{1}
\end{equation}
Dotting with $\bfB=\curl{\bf A}$ we have 
\begin{equation}
\bfE\cdot \bfB=- \nabla \Phi \cdot \bfB 
-(1/c){\bf B}\cdot \partial_t{\bf A}.
\label{02}
\ee
After straightforward algebraic manipulation, application of Maxwell's
equations  
and $\nabla\cdot \bfB=0$, this equation implies
\beq
\bfE\cdot \bfB
=- (1/2)\nabla\cdot   \Phi\bfB 
+(1/2)\nabla \cdot({\bf A}\ts {\bf E}) \nn\\
-(1/2c)\partial_t( 
{\bf A}\cdot \bfB) 
=(-1/2c)\partial_\mu{H}^{\mu}(\bfB)=\eta\bfJ\cdot\bfB,
\label{03a1}
\eeq
where 
\beq
H^{\mu}(\bfB)=(H_{0},H_i)
=[{\bf A}\cdot \bfB, c\Phi\bfB 
-c{\bf A}\ts {\bf E}]
\label{heldef}
\eeq
is the magnetic helicity density 4-vector [24],
and the contraction has been done with the 4 x 4 matrix $\eta_{\mu\nu}$ where
$\eta_{\mu\nu}=0$ for $\mu \ne \nu$, $\eta_{\mu \nu}=1$ for $\mu=\nu=0$ and 
$\eta_{\mu\nu}=-1$ for $\mu=\nu >0$.
Taking the average of (\ref{03a1}) gives
\be
\partial_\mu{\overline H}^{\mu}({\bfB})=-2c\lb\bfE\cdot \bfB\rb=
-2c\bbE\cdot \bbB-2c\lb\bfe\cdot \bfb\rb=-2c\eta\lb\bfJ\cdot\bfB\rb
\label{03a2}.
\ee

If, instead of starting with the total $\bfE$ as in (\ref{1}), 
we start with $\bfe$ and then dot with $\bfb$ and average, 
the analagous derivation replaces (\ref{03a2}) by 
\be
\partial_\nu {\overline H}^{\mu}(\bfb)=
-2c \lb {\bfe}\cdot {\bfb}\rb,
\label{14}
\ee
where ${\overline H}^{\mu}(\bfb)$ indicates the average of 
${H}^{\mu}(\bfb)$. The latter is defined like (\ref{heldef}) but with
the corresponding fluctuating quantities replacing the total quantities.
Similarly, starting with $\bbE$ and dotting with $\bbB$, 
gives
\be
\partial_\mu H^\mu(\bbB) 
=-2c {\overline {\bf E}}\cdot {\bbB},
\ee
where ${H}^{\mu}(\bbB)$ is defined as in (\ref{heldef}) but with
the corresponding mean quantities replacing the total quantities.
Now consider two cases.

\bigskip
\hskip0.5in{ \sl  1. Case 1}
\bigskip
 
In this case we consider statistically stationary turbulence 
in which the scale of averaging is equal to the universal scale, or 
equivalently for present purposes, that 
the overbar indicates averaging over periodic boundaries.  
In this case, the spatial divergence terms on the left of
(\ref{03a2}) become surface integrals and vanish. 
Rewriting this  (\ref{14}), and discarding the divergenece
terms gives
\be
\lb {\bfe}\cdot {\bfb}\rb=-{1\over 2c}\partial_t \lb{\bfa}\cdot {\bfb} \rb
\label{17}
\ee
Thus, eqn (\ref{mon23}) is resistively limited  unless 
significant non-stationarity is allowed.

Let us apply this to the specific case of statisically
stationary turbulence in which a uniform 
mean field is imposed over a periodic box, and for which 
the averaging scale is the scale of the box.  
Then the mean field cannot change with time, and has no gradients.
This is the case of Ref [18].
Then the right side of (\ref{17}) vanishes because
not only are zeroth order turbulent correlations 
(those computed to zeroth order in the mean field) 
stationary, but all higher order corrections must also 
be stationary.  The mean field does not change with time so  
no mean quantity varys on macroscopic time scales. 
Incidentally, when expanded in powers of the mean field, the lowest
order contribution to (\ref{17}) enters at second order [23].

For the uniform field in a periodic box, (\ref{mon23}) then implies
that 
\be
 -c^{-1}\lb{\bfv\times \bfb}\rb \cdot \bbB 
=\alpha {\bbB}^2/c =  c^{-1}\la \lra{\bfb\cdot\nb\times \bfb },
\lab{mon23q}
\ee
where $\alpha=\alpha_{33}$ for a uniform field in the $z$ direction.
Rearranging  we have
\beq
|\alpha| = \left|{ \la \lra{\bfb\cdot\nb\times \bfb }\over  \bbB^2}\right|.
\label{19}
\eeq
Now if we assume that $b$ and $v$ follow e.g. a $k^{-5/3}$ Kolmogorov energy 
spectrum,  we then have 
\beq
|\alpha| \lsim \left|{{k_0 b_0^2} \tau_{0}\over R_M^{3/4} \bbB^2/v^2}\right|
\sim \left|{\alpha_0\over R_M^{n} \bbB^2/b_0)^2}\right|
\label{answer}
\eeq
where $n=3/4$  if the current helicity is dominated by large wavenumber
and $n=1$ if it is dominated by small wavenumber.
The  $v_0,b_0$ are the speed and 
fluctuating field magnetic energy 
of the dominant energy containg eddies, $k_0$
is the wavenumber for that scale, 
$\tau_0$ is the associated eddy turnover time
and the magnetic Reynolds number 
$R_M=v L/\la$, where $L$ is the scale of the energy dominating
eddies.  The latter similarity in (\ref{answer}) 
follows for near equipartition between magnetic and kinetic energies
on the outer scale, and we have written the kinematic $\alpha$ as 
$\alpha_0$.   (Note that the ``Pouquet correction'' [21]
to $\alpha$ which is $\propto \lb\bfb\cdot\curl \bfb\rb$ enters to the same
order in the mean field as the standard $\lb\bfv\cdot\curl\bfv\rb$ term [10,32],
so we consider either to be representative of the lowest order
contribution).
The value of $\alpha$ in this
case is ``resistively limited,'' however the Reynolds number factor
on the bottom does not represent a dynamical backreaction.
It is an a priori implication of the imposed boundary conditions.
Thus simulations which invoke periodic boundary conditions 
and find this level of suppression [18], may not be seeing the
effect of backreaction dynamics, but of the boundary conditions.

\bigskip
\hskip0.5in {\sl 2. Case 2}
\bigskip

In this case, consider 
the system (e.g. Galaxy or Sun)  volume V $<<$  universal volume.
Integrating (\ref{03a1}) over all of space, $U$, then gives 
\beq
\int_U \bfE\cdot \bfB\ d^3x=- (1/2)\int_U\nabla\cdot   \Phi\bfB\ d^3x 
+(1/2)\int_U\nabla\cdot( {\bf A}\ts {\bf E})\ d^3x \nn\\
-(1/2c)\partial_t\int_U{\bf A}\cdot \bfB\ d^3x 
=-(1/2c)\partial_t{\EuScript H}(\bfB)=\int_U \eta {\bf J}\cdot \bfB d^3x  ,
\label{3a1}
\eeq
where the divergence integrals vanish when converted to surface integrals at infinity.
We have defined the global magnetic helicity  
\begin{equation} 
{\EuScript H}(\bfB)
\equiv\int_U{\bf A}\cdot \bfB\ d^3x,
\label{3aa}
\end{equation}
where  $U$ allows for scales much larger than the mean field scales.
It is straightforward to show that a parallel argument 
for the mean and fluctuating fields respectively leads to 
\be
\partial_t{\EuScript H}(\bbB)=
\partial_t\int_U\bbA\cdot\bbB\ d^3x
=-2c\int_U\bbE\cdot\bbB\ d^3x
\label{3aaa}
\ee
and
\be
\partial_t{\overline{\EuScript H}}(\bfb)=
\partial_t\int_U
\lb{\bf a}\cdot{\bf b}\rb\ d^3x
=-2c\int_U\lb\bfe\cdot\bfb\rb\ d^3x=-2c\int_U\bfe\cdot\bfb\ d^3x
=\partial_t{{\EuScript H}}(\bfb),
\label{3aab}
\ee
where the penultimate equality in (\ref{3aab}) follows from 
redundancy of averages.


We now split  (\ref{3aaa}) and (\ref{3aab})  
into contributions from inside and outside the rotator.
One must exercise caution in doing so because ${\EuScript H}$
is gauge invariant and physically meaningful only 
if the volume $U$ over which ${\EuScript H}$ is integrated is bounded by 
a magnetic surface (i.e. normal component of $\bfB$ vanishes at the surface)
, whereas the surface separating the 
outside from the inside of the rotator is not magnetic in general.

Ref [25] shows how to construct a revised gauge invariant quantity  
called the relative magnetic helicity.  
This can be written as 
\beq
{\EuScript H}_{R,i}({\bf B}_i)
= {\EuScript H}({\bfB}_i,{\bf P}_o)
-{\EuScript H}
({\bf P}_i,{\bf P}_o)  
\label{relhel}
\eeq
where the two arguments represent inside and outside the body respectively,
and $\bf P$ indicates a potential field.   
The relative helicity of the inside region is thus 
the difference between the actual helicity and the helicity
associated with a potential field inside that boundary. The use of
${\bf P}_i$ is not arbitrary in (\ref{relhel}), 
and is in fact the field configuration of lowest energy.  
While (\ref{relhel})
is insensitive to the choice of external field [25], 
it is most convenient to take it to be a potential field as is 
done in (\ref{relhel}) symbolized by ${\bf P}_o$. 
The relative helicity of the outer region, ${\EuScript H}_{R,o}$,
is of the form (\ref{relhel}) but with the $o$'s and $i$'s reversed.
The ${\EuScript H}_R$
is invariant even if the boundary is not a magnetic surface.

The total global helicity, in a  
magnetically bounded volume divided into the sum of internal and 
external regions, $U=U_{i}+U_{e}$, satisfies  [25]
\be
{\EuScript H}(\bfB)={\EuScript H}_{R,o}(\bfB)+{\EuScript H}_{R,i}(\bfB),
\label{r4aaa}
\ee
when the boundary surfaces are planar or spherical.
This latter statement on the boundaries means the
vanishing of an additional term associated with potential fields
that would appear in (\ref{r4aaa}).
Similar equations apply for $\bbB$ and $\bfb$, so (\ref{3aaa})
and (\ref{3aab}) can be written
\be
\partial_t{{\EuScript H}}(\bbB)
= \partial_t{{\EuScript H}}_{R,o}(\bbB) +\partial_t{{\EuScript H}}_{R,i}(\bbB),
\label{r4aab1}
\ee
and 
\be
\partial_t{{\EuScript H}}(\bfb)
=\partial_t{{\EuScript H}}_{R,o}(\bfb)
 +\partial_t{{\EuScript H}}_{R,i}(\bfb)
\label{r4aab2}
\ee
respectively. According to equation (62) of Ref [25], 
\be
\partial_t{{\EuScript H}}_{R,i}(\bfB)=
-2c\int_{U_{i}}{\bfE\cdot\bfB} d^3x+2c\int_{D U_i}({\bfA}_p\ts {\bfE})\cdot d{\bf S},
\label{r4aac}
\ee
where ${\bfA}_p$ is the vector potential corresponding to a potential field ${\bf P}$ in $U_e$, and $DU_i$ indicates integration on the boundary surface of
the rotator. 
Similarly, we have 
\be
\partial_t {{\EuScript H}}_{R,i}(\bbB)=
-2c\int_{U_{i}}{\bbE\cdot\bbB} d^3x
+2c\int_{D U_i}({\bbA}_p\ts 
{\bbE})\cdot d{\bf S}
\label{r4aad}
\ee
and
\be
\partial_t{{\EuScript H}}_{R,i}(\bfb)=
-2c\int_{U_{i}}{\bfe\cdot\bfb} d^3x+2c\int_{D U_i}{\bf a}_p\ts 
{\bfe}\cdot d{\bf S}.
\label{r4aae}
\ee
Note again that the above internal relative helicity time derivatives are 
both gauge invariant and independent of the field assumed in the
external region.  If we were considering the relative helicity of the
external region, that would be independent of the actual field in the internal
region.

Now if we take the averaging scale to be less than or equal to 
the $U_i$ scale, we can then
replace 
(\ref{r4aae}) by
\be
\lb{\bfe\cdot\bfb}\rb
=-{1\over 2c}
\partial_t{{\EuScript H}}_{R,i}(\bfb)+\lb\div({\bf a}_p\ts 
{\bfe})\rb,
\label{r4aaeb}
\ee
where the brackets indicate integrating over $U_i$ or smaller.
We now see that even if the first term on the right of (\ref{r4aaeb}) vanishes,
$\lb{\bfe\cdot\bfb}\rb$ contributes a surface term to 
(\ref{mon23}) that need not vanish.  The turbulent
EMF is not resistively limited as in the case of the previous section.
The surface term can strongly dominate the resistive contribution.
Thus in a steady state for $R_M>>1$, an outflow of magnetic helicity 
is likely essential to keeping the dynamo operating.
For the Sun, the steady state would refer to time scales longer than
the longest eddy turnover time, but shorter than the predicted 
11 year Solar cycle.
Note that  the right of (\ref{03a2}) 
is small for  
for large magnetic Reynolds numbers, so the flux of helicity has contributions
from the small and large scale field.
If the surface term vanishes,  helicity could instead be 
``injected''  through the time derivative term of $(\ref{03a2})$.
This has known  application to  Tokomaks [26].


In short, $steady$ dynamo action unrestricted by 
resistivity is possible in case 2 but not in case 1.
This is consistent with current simulations.
Case 1 is represented by  
Refs [21] and [22], which find that when the boundary
conditions are periodic, dynamo action proceeds at resistively limited rates.
In [27] action seems to proceed more rapidly, and this corresponds to a 
case 2 simulation, though the resolution and $R_M$ were small.


\bigskip
\ni {\bf III.  ENERGY FLOW TO CORONAE}
\bigskip

Here we explore only the deposition of relative helicity to the
exterior and the associated total magnetic energy
without addressing how the energy converts to particles or flows.
We assume that the rotator is in a steady state
over the time scale of interest, so the left sides of 
(\ref{r4aad}) and (\ref{r4aae}) vanish. 
Note that in a system like
the Sun where the mean field flips sign 
every $\sim 11$ years, the steady state is
relevant for time scales less than this period,
but greater than the eddy turnover time ($\sim 5 \ts 10^4$ sec).
Beyond the $\sim 11$ year times scales, the mean large and small
scale relative helicity contributions need not separately
be steady and the left hand sides need not vanish.

The helicity supply rate, represented by the volume integrals
(second terms of (\ref{r4aad}) and (\ref{r4aae})),
are then equal to the integrated flux of relative magnetic 
helicity through the surface of the rotator.  Moreover, from 
(\ref{03a2}), we see that the integrated flux of the large 
scale relative helicity, $\equiv {\EuScript F}_{R,i}(\bbB)$,
and the integrated flux of small scale relative helicity,
$\equiv {\EuScript F}_{R,i}(\bfb)$,
are equal and opposite.  We thus have
\be
{\EuScript F}_{R,i}(\bbB)=
-2c\int_{U_{i}}{\bbE\cdot\bbB} d^3x
=2c\int_{U_{i}}{\bfe\cdot\bfb} d^3x=
-{\EuScript F}_{R,i}(\bfb).
\label{r4aaf}
\ee
To evaluate this,  we use  (\ref{2.4a}) and (\ref{2.4aa}) to
find
\be
\bbE=-c^{-1}(\alpha\bbB-\beta\curl\bbB),
\label{r4aaf2}
\ee
throughout $U_i$.  
Thus  
\be
{\EuScript F}_{R,i}(\bbB)= -{\EuScript F}_{R,i}(\bfb)=
2 \int_{U_i}(\alpha\bbB^2-\beta\bbB\cdot\curl\bbB)d^3{\bf x}.
\label{r4aag}
\ee
This shows the relation between the 
equal and opposite large and small scale relative helicity
deposition rates and the dynamo coefficients.

Now the realizability of a helical magnetic
field requires its turbulent energy spectrum, $E^M_k$, to satisfy [28]
\be
E^M_k(\bfb)\ge {1\over 8\pi} k|{{\EuScript H}}_k
(\bfb)|\; , \lab{10}
\ee
where ${{\EuScript H}}_k$ is the magnetic helicity at wavenumber $k$.
The same argument also applies to the mean field energy spectrum, so that 
\be
E^M_k(\bbB)\ge {1\over 8\pi} k|{{\EuScript H}}_k(\bbB)|. 
\lab{10a}
\ee
If we assume that the time and spatial dependences are separable
in both $E^M$ and ${{\EuScript H}}$, then 
a minimum power delivered to the corona can be derived.
For  the contribution from the small scale field, 
we have 
\beq
\begin{array}{r}
\dot E^M(\bfb) = 
\int {\dot E}^M_k(\bfb) dk \ge {1\over 8\pi}\int k|
{\EuScript F}_{k,R,i}(\bfb)| \, dk
\ge  {k_{\rm min}\over 8\pi} \int | {{\EuScript F}}_{k,R,i}(\bfb)| \, dk
\nn\\ 
\ge {k_{\rm min}\over 8\pi} |{\EuScript F}_{R,i}({\bfb})|
={k_{\rm min}\over 8\pi} | {\EuScript F}_{R,i}(\bbB)|,
\end{array}
\lab{11}
\eeq  
where the last equality follows from the first equation in 
(\ref{r4aag}).
The last quantity is exactly the lower limit on 
$\dot E^M(\bbB)$.  Thus the sum of the lower limits on the total power 
delivered from large and small scales is 
${k_{\rm min}\over 8\pi} | {\EuScript F}_{R,i}(\bbB)|
+{k_{\rm min}\over 8\pi} | {\EuScript F}_{R,i}(\bfb)|
=2{k_{\rm min}\over 8\pi} | {{\EuScript F}}(\bbB)|$.  
Now for a mode to fit in the rotator, $k>k_{\rm min}=2\pi/h$, 
where $h$ is a characteristic scale height of the turbulent layer.  
Using (\ref{r4aag}), the total estimated energy delivered to the corona
(=the sum of the equal small and large scale contributions) is then
\beq
\dot E^M  \ge 2{k_{\rm min}\over 8\pi} |{\EuScript F}_{R,i}(\bfb)|
=2{k_{\rm min}\over 8\pi} |{\EuScript F}_{R,i}(\bbB)|
={{\rm V} \over h} \left| {\alpha\ob^2-\beta\bbB\cdot\curl\bbB}\right|_{ave},
\label{result}
\eeq
where $\rm V$ is the volume of the turbulent rotator and $ave$ indicates
volume averaged. We will 
assume that the two terms on the right of (\ref{result}) 
do not cancel, and use the first term of (\ref{result}) as representative.

\medskip


Working in this allowed time range for the Sun $(5\ts 10^{4}{\rm sec}<t< 11
{\rm yr}$ we  apply 
(\ref{result}) 
to each hemisphere of the Sun, and using the first term as an
order of magnitude estimate, obtain
\beq
{\dot E}^M \gsim  \left({2\pi R_\odot^2 \over 3}\right)\alpha \bbB^2=
10^{28} \left({R \over 7\ts 10^{10}{\rm cm}}\right)^2 
\left({\alpha \over 40{\rm cm/s}}\right)
\left({{\overline B} \over  150 {\rm G}}\right)^2 {\rm {erg\over s}}
\label{sun}
\eeq
where we have taken $\alpha\sim 40\,$cm~s$^{-1}$ ,
and we have presumed a field of 150G 
at a depth of $10^4$km beneath the solar surface in the convection zone,
which is in energy  equipartition with turbulent kinetic motions [2].

As this energy deposition rate is available for 
reconnection which can generate Alfv\'en waves, drive winds, and 
energize particles, we must compare this limit with the total of
downward heat conduction loss, radiative loss, and solar wind energy flux
in coronal holes, which cover $\sim 1/2$ the area of the Sun.  
According to [29] this amounts to an approximately
steady activity of $2.5\times
10^{28}$erg~s$^{-1}$, about 3 times the predicted value of (\ref{sun}).
Other supporting evidence  
for deposition of magnetic energy and magnetic + current helicity [30] 
in the Sun is discussed in Ref [31].

AGN and the Galactic ISM represent other likely sites of mean field dynamos.
Ref [31] discusses the associated energy
deposition to the coronae of these systems as well.
For the Galaxy,  ${\dot E}^M \gsim ({\pi R^2})\alpha \ob^2
\sim10^{40} ({R / 12{\rm kpc}})^2$ $\ts ({\alpha / 10^5{\rm cm/s}})
({\ob/ 5\ts 10^{-6}{\rm G}})^2 {\rm erg/s}$
in each hemisphere.
This is consistent with coronal energy input rates required by [32]
and [33].
For AGN accretion disks, the deposition rate seems to be consistent
with what is required from X-ray observations.
Independent of the above, 
the most successful paradigm for X-ray luminosity in AGN 
is coronal dissipation of magnetic energy  [34].

\bigskip
\ni {\bf  IV. DISCUSSION AND OPEN QUESTIONS}

\bigskip
{\bf A.  Role of boundary diffusion and winds}
\bigskip

We have seen that a steady MFD unlimited by resistivity 
requires the helicity to flow through the rotator boundary.
However, merely open boundary conditions may not be enough:
some mechanism must sustain the diffusion
of the mean field at the boundary. Near the surface, 
buoyancy or winds [35] 
and winds may have to play a role.  
Note however, that it is the mean field which needs to diffuse, not 
necessarily the total field, and not the matter [5].  

For the Galaxy, turbulent diffusion of the mean
magnetic field across the boundary 
is also required to maintain a quadrupole geometry with a net
flux inside the disk. 
Ref [36] argues that  surface diffusion of total magnetic field
for the Galaxy is  difficult.  Thus the diffusion 
of the mean and/or total field out of a galaxy remains an open question.
For the Sun, the solar cycle also requires 
diffusion of mean field through the boundary. 
The flow of helicity 
would appeal to the same dynamics needed by these constraints.

The role of boundary diffusion is essential for understanding
whether the present Galactic field is primordial or produced in situ
and how the solar dynamo accounts for the surface fields.

\bigskip
{\bf B. Choosing simulation boundary conditions: transient vs steady phases}
\bigskip

Ideally, the best simulation to test the dynamo would
be a fully global simulation with significant scale separation
between the system size and the turbulent scale, and would properly
include the boundary diffusion mechanisms with open boundary conditions.
However it may be possible to invoke periodic boundary conditions to simulate
rapid dynamo action in the following ways.  The
first represent transient phases, the last approaches represent
steady action.

A first transient phase during which  
rapid dynamo action should incur, is the time period during which the
helicity on the small scale builds up.  During this period
the dynamo can grow rapidly.  After this period one is subject
to the boundary conditions. This has been investigated numerically
in Ref [37], where the dynamo $\alpha$ is shown not to be significantly
quenched in this early phase.

A second time scale over which rapid dynamo action could 
be seen in a periodic box, even after the small scale helicity
saturates, would be facilitated by 
significant scale separation between the box size and the 
scale of the input turbulent scale.
One could inject kinetic helicity only into a sub-volume
of the box and observe the inverse cascade of magnetic helicity and
magnetic energy in this sub-volume before the material 
diffuses all the way across the box. During this transient period,
even if the small scale helicity saturates, the 
surface terms can play an important role.  The length of this
period of course depends on the extent of scale separation.

Finally, there may be a way to use a periodic box to actually
get steady rapid dynamo action.
One can inject helicity inside the sub-volume described above,
and invoke a one-way diffusive valve out of this region.
The helicity of one sign should grow inside the forced region,
while the opposite sign would be shed to the exterior region.
One could then surround the box with a resistive layer so
that this external helicity dissipates rather than feeds back
into into the system.

\bigskip
{\bf C. Previous analytical and numerical suppression results}
\bigskip

To date, analytic and numerical studies that 
suggest resistively limited dynamo action
either (1) invoke periodic boundary conditions (as described above), and/or 
(2) are 2-D, or (3) do not distinguish between zeroth order isotropic
components of the turbulence and the higher order anisotropic
perturbations for a weak mean field [31].
This existence of an alternative explanation for
the catastrophic suppression makes the computed
suppression ambiguous.  
This does not mean that the physical concepts
found in the strong suppression results are invalid, but just that they 
may be valid only for the restricted cases considered.  

The limitations can be specifically pinpointed. 
For example, [38] derives the now
famous ``Zeldovich relation''  
$\lb b^2\rb/\bB^2\sim R_M$. But this relation 
is for $2-D$, ignores boundary terms, and is non-trivial only when
the magnetic energy is dominated on small scales (unlike
3-D simulation results).

The Zeldovich relation arises from first deriving, in the absence of boundary
terms, the evolution equation for the average of the
square of the vector potential from an initially  uniform magnetic
field in 2-D.
We choose the vector potential to be in the $z$ direction, and 
thus we have $A=A_z$.  The resulting evolution equation is 
\beq
\partial_t\lb{A_z}^2\rb=
-2\lambda\lb(\nabla{A_z})^2\rb.
\label{z571}
\eeq
Note that there is only a single term on the right hand side
because of the boundary conditions.  
We now appeal to the positive definite nature of the last term
to rewrite the equation
\beq
\partial_t\lb{ A_z}^2\rb=-2\lambda\lb{A_z}^2\rb/\delta^2(t),
\label{z572}
\eeq
where $\delta(t)$ represents the dominant scale of the
turbulent field. 
In 2-D, when surface terms are ignored,  the mean field cannot change. However
the small scale field can grow in response to the turbulent stretching.
In analytic studies [eg. 16] and numerical simulations [e.g. 37] (which
happens to be 3-D) this type of behavior can be described 
by a decrease of $\delta(t)$, that is  $d\delta(t)/d t<0$,
as the turbulent eddies cascade to smaller and smaller scales.
Equipartition between magnetic and kinetic energy is 
first reached on the dissipative scale as there the eddy turnover time
is the shortest.  When the field peaks at this scale, the
dissipation rate in (\ref{z572}) is maximized because $\delta(t)$ reaches
its minimum. The end of kinematic regime occurs when the field
saturates at the resistive scale.   In fact the {\it maximum} dissipation rate 
occurs if  $R_M \sim b^2/\bB^2$.
(Note however, that if $R_M >> b^2/\bB^2$, 
saturation at $b^2=v^2$ would occur before $R_M$ enters the relation.)

Soon after the maximum dissipation is achieved, 
$d\delta/dt >0$. That is, the scale $\delta(t)$  
increases as the dynamic regime sets in and equipartition is reached
on successively larger and larger scales.
The dissipation rate then decreases in the non-linear
dynamic regime. This is the  ``suppression'' of dissipation
that is directly implied by (\ref{z572}).
We refer to this as ``dissipation'' rather than ``diffusion''
because the only term determining the time evolution of $\lb A_z^2 \rb$
has an explcit $\la$ as a coefficient.  
 
In fact, the original Zeldovich result was meant to be a limitation
on the growth of $\lb b^2\rb$ from a fixed $\bB$,  
the latter of which was restricted not to change due to 2-D calculation 
and the choice of boundary conditions.  Nothing was said about the
turbulent diffusion of field lines or the ability of the mean field to grow.
Zeldovich recognized that the 2-D case did not necessarily imply  
much about the 3-D case, the case when boundary terms are included, or the
case when $\bB$ is allowed to grow.  

That being said, there are some important related issues to understand in
future work.
The observation that at least the 2-D Lagrangian chaos properties of the flow
seem to change in the presence of a weak mean field for turbulence in a 
periodic box [39] needs to be understood in relation to 
the imposed boundary conditions and the
shape of the magnetic energy spectrum.  
The connection between this observed restricted 2-D diffusion and the
Zeldovich relation is actually subtle, and 
also requires further investigation.
One point to  consider in this context is 
that when the diffusion time across the box is comparable to
the Alfv\'en crossing time across the box, numerical
effects may be playing a strong 
role in determining the results and must be kept in mind.


At present there is a dearth of simulations on
3-D turbulent diffusion of the mean field.  However molecular
cloud simulations [40] which do not explicitly consider an application
to dynamo theory, can actually be interpreted to imply the absence of 
catastrophic quenching of mean field diffusion.

Finally, we note that Zeldovich et al. produced a 
second approach to this relation between the mean field strength
and $R_M$ which was thought to be applicable to 
3-D [4]. In 3-D,  a logarithmic dependence on $R_M$ arises, while 
the method also seems to reproduce the 2-D result in the 2-D limit. 
However, this approach may be flawed as it does not distinguish
between isotropic and anisotropic components of the turbulence.
This will be discussed further elsewhere.

\bigskip
{\bf D. Coronal Activity}
\bigskip

The estimated energy deposition rates 
are consistent with the coronal  + wind
power from the Sun, Galaxy and Seyfert Is [31].
The helical properties also seem to agree well in the solar case where they
can be observed [30].
The steady flow of magnetic energy into  coronae thus provides an interesting
connection between mean field dynamos and coronal dissipation paradigms
in a range of sources.  A reasonably steady (over time scales
long compared to turbulent turnover time scales), active corona with
multi-scale helical structures, provides a self-consistency check for 
a dynamo production of magnetic field
in which there is exponential field  growth inside the body on
small and large scales. If the growth rate  
were only linear, the corona and wind output might be 
more episodic with fluctutuations not by many orders of magnitude
but by an order of magnitude or less.
The latter seems to be consistent with observations of the Sun
and coronae of Seyfert AGN [41], suggesting exponential field
growth inside the rotator.


\bigskip
{\bf E. Role of magnetic Prandtl number}
\bigskip

Much of the numerical and analytic work to date with respect to the mean
field dynamo problem has focused on unit Prandtl number $Pr\equiv 
\la/\eta$ (the ratio of
the magnetic to kinetic viscosities.)  In nature, this number is rarely
near unity. In the Sun $Pr << 1$, whereas in the Galaxy
$Pr >> 1$.  There are currently several numerical [19,42,43] and analytic 
studies [43]
addressing the role of the magnetic Prandtl number.
For unit Prandtl number, the magnetic and kinetic energies
approach equipartition from the the input scale down to the dissipation scale.
For $Pr>>1$, Ref [41] finds that the magnetic energy dominates
the turbulent energy at scales below the viscous cutoff scale,
and that the magnetic energy initially peaks on these tiny scales.
With time, the peak seems to migrate back toward the input scale, 
although runs have not been done long enough to see what the saturated
state is like.  These simulations do not have a large enough
inertial range, and so are far from being able to assess the power build up
at scales even larger than the forcing scales, as required to test the MFD.
nor do they consider helical forcing.

Ref [21] finds that the non-local inverse
cascade is not much affected by the large $Pr$ when compared to the small
$Pr$, but intially sees a similar development of the small scale magnetic energy
as in Ref [42]. The dynamic range is also limited.  Refs 
[43] appears consistent with this. The work also 
suggests that the contribution to $\alpha$ from the current
helicity might exceed that from the kinetic helicity for large 
Prandtl number.  More work is needed to understand the effect
of large and small Prandtl numbers.

\bigskip
\ni {\bf V. CONCLUSIONS}
\bigskip

We have emphasized that the 
MFD represents  a framework for understanding an inverse 
cascade of magnetic helicity:  Kinetic helicity imposed 
at small scales pumps magnetic helicity from small to large scales 
through an inverse cascade [21,22,44].  This process 
seems to be non-local [22] in that helicity ``jumps'' 
directly from small to large scales. Accompanying the magnetic helicity  is 
a growth of magnetic energy.  

We suggest that the MFD 
can in principle proceed much more rapidly for real astrophysical rotators,
when compared to simulations in periodic boxes 
in which the turbulence is homogeneous.
For such simulations, surface terms are
ignored and so the  magnetic helicity 
is nearly conserved for large magnetic Reynolds numbers.
Generating magnetic helicity
and magnetic energy on the large scales from an MFD then requires 
generating a compensating magnetic helicity at or below the input scale. 
In a steady state, the small scale helicity would drain 
only through resistive dissipation.  In this case the growth rate of the large
scale field is resistively limited.

But in real astrophysical systems, the boundary terms relax
the helicity conservation constraint. The growth rate 
of the large scale helicity (and thus large scale field) 
is then limited by the rate at which the compensating helicity flows
out the boundary. This rate can in principle be much faster than the 
resistively limited rate.  The message is that the boundary
terms are likely important for astrophysical MFD's and so backreaction 
studies 
which show suppression but do not allow boundary terms leave
the ambiguity as to whether the suppression is actually due to the dynamic
backreaction or is simply due to the boundary conditions.

Note that the role of shear for the generation of magnetic

A steady flow of
magnetic helicity into the corona is expected for 
an astrophysical rotator 
harboring a vigorous $\alpha-\Omega$ mean field dynamo.
The helicity escape rate leads to
a lower limit on the total magnetic energy deposition into the corona.
When the corona itself is turbulent, there should also 
be an inverse cascade of magnetic helicity 
in any wind driven outward, thus the dominant magnetic helicity scale would 
appear to increase on increasingly large distances from the source.

In addition to the conclusions above, we 
discussed a number of open issues: the choice of boundary
conditions for simulations, the need to understand the boundary
diffusion in real systems, some implications of previous analytic
work, the predicted variability for a corona, and the role of
the magnetic Prandtl number.

\medskip


\centerline {\bf Acknowledgements}
E.B. acknowledges support from DOE grant DE-FG02-00ER54600
and thanks the APS/DPP Quebec 2000 meeting organizers for the 
invitation.

\centerline{\bf References}

\def\item{\ni}

\ni $^1$ 
H.K. Moffatt, H. K. {\sl Magnetic
Field Generation in Electrically Conducting Fluids}, (Cambridge:
Cambridge University Press, 1978)

\ni  $^2$ 
E.N. Parker, {\it Cosmical Magnetic Fields} (Oxford: Clarendon
Press, 1979).

\noindent $^3$  F. Krause \& K.-H. R\"adler 
{\it 
Mean-field magnetohydrodynamics and dynamo theory}, (New York: Pergamon, 1980).

\ni $^4$  
YA. B. Zeldovich , A.A. Ruzmaikin, , 
and D.D. Sokoloff, {\sl Magnetic Fields in Astrophysics
}, (New York: Gordon and Breach, 1983).

\ni $^5$  A.A. Ruzmaikin,  
A.M. Shukurov, D.D. Sokoloff,  {\sl Magnetic Fields of
Galaxies}, (Dordrecht: Kluver Press, 1988).

\ni $^6$  R. Beck et al.,  Ann. 
Rev.  Astron. Astrophys., {\bf 34}, 155 (1996).

\ni $^7$  M. Steenbeck, F. Krause, \& K.-H. R\"adler, Z. Naturforsch.
{\bf 21a}, 369 (1966).

\ni $^8$  E.N. Parker, ApJ {\bf 122}, 293 (1955).

\ni $^9$  E.G. Blackman \& T. Chou , ApJ, {\bf 489},
L95 (1997).

\ni $^{10}$   
G.B. Field, E.G. Blackman,  \& H. Chou,  ApJ {\bf 513}, 638 (1999).

\ni  $^{11}$ A.V. Gruzinov \&  P.H. Diamond P.H., PRL, 72, 1651 (1994);
A.V. Gruzinov \& P.H. Diamond, Physics of Plasmas, {\bf 2} 1941 (1995);
A. Bhattacharjee \& Y. Yuan, ApJ, {\bf 449} 739 (1995).

\ni $^{12}$ A.V. Gruzinov \& P.H. Diamond, Physics of Plasmas, {\bf 3} 1853 (1996). 

\ni $^{13}$ E.G. Blackman \& G.B. Field, ApJ {\bf 521} 597 (1999).

\ni $^{14}$    T.G. Cowling,  {\sl Magnetohydrodynamics},
(New York: Interscience, 1957).

\ni $^{15}$   J.H. Piddington,  {\it
Cosmical Electrodynamics}, (Malbar: Krieger. 1981).

\ni $^{16}$   R.M. Kulsrud \& S.W. Anderson, ApJ {\bf 396} 606 (1992).

\ni $^{17}$   L.L. Kitchatinov,
V.V. Pipin,  G. R{\"u}diger, G., \& M. Kuker, Astron. Nachr., 
{\bf 315}, 157 (1994).

\ni $^{18}$ F. Cattaneo, \& D.W. Hughes, Phys. Rev. E. {\bf 54}, 4532 (1996).

\ni $^{19}$    S. Vainshtein, PRL, {\bf 80}, 4879 (1998).

\ni $^{20}$   J.A. Markiel \& J.H. Thomas, ApJ, {\bf 523} 827 (1999)

\ni $^{21}$ A. Pouquet, U. Frisch, \& J. Leorat,  JFM {\bf 77}, 321 (1976).

\noindent $^{22}$ A. Brandenburg A.,  submitted to ApJ, astro-ph/0006186 
(2000).

\noindent $^{23}$ E.G. Blackman,  \& G.B. Field, ApJ {\bf 534} 984 (2000a).

\ni $^{24}$ G.B. Field  
in {\it 
Magnetospheric Phenomena in Astrophysics}, R. Epstein \& W. Feldman, eds.\
AIP Conference Proceedings 144 (Los Alamos: Los Alamos Scientific
Laboratory 1986), p324.

\ni $^{25}$ M.A. Berger  \& G.B. Field,  JFM, {\bf 147} 133 (1984).

\ni $^{26}$ H. Ji, Physical Review  Letters, {\bf 83} 3198 (1999);
P.H. Diamond, personal communication.

\ni $^{27}$  A. Brandenburg \& K.J. Donner, 
MNRAS {\bf 288}, L29 (1996).

\ni $^{28}$ U. Frisch, A. Pouquet, J. L\'eorat \& A. Mazure,  JFM {\bf
68}, 769 (1975).

\ni $^{29}$ G.L. Withbroe, \& R.W. Noyes, Ann. Rev. Astron.
Astrophys. {\bf 15}, 363 (1977); 

\ni $^{30}$ D.M. Rust  \& A. Kumar , ApJ, {\bf 464}, L199 (1994).


\noindent $^{31}$ 
E.G. Blackman,  \& G.B. Field, MNRAS, in press,
astro-ph/9912459 (2000b).

\noindent $^{32}$ 
B.D. Savage,  in {\it The Physics of the Interstellar Medium
and Intergalactic Medium},  A. Ferrara, C.F. McKee, C. Heiles, \& 
P.R. Shapiro eds. ASP conf ser vol 60. (San Francisco:  PASP, 1995) p233.

\ni $^{33}$ R.J. Reynolds, L.M. Haffner, S.L. Tufte , ApJ, {\bf 525}, L21 
(1999).

\ni $^{34}$ F. Haardt  \& L. Maraschi, ApJ, {\bf 413}, 507 (1993);
G.B. Field \& R.D. Rogers., ApJ, {\bf 403}, 94 (1993); 
T. DiMatteo E.G. Blackman \& A.C. Fabian, MNRAS, {\bf 291} L23 (1997);
A. Merloni \& A.C. Fabian, accepted to MNRAS, astro-ph/0009498 (2000);
\ni  K.A. Miller \& J.M. Stone, ApJ {\bf 534} 398 (2000).

\ni $^{35}$ D. Moss, A. Shukurov \& D.  Sokoloff, A\&A {\bf 340}, 120 (1999);
A. Bardou, B.V. Rekowski, W. Dobler, A. Brandenburg, \& A. Shukurov,
submitted to A\& A, astro-ph/0011545.

\ni $^{36}$ R.R. Rafikov\& R.M. Kulsrud, MNRAS, {\bf 314}, 839 (2000). 

\ni $^{37}$ H. Chou, submitted to ApJ, astro-ph/0011548 (2000)

\ni $^{38}$ Ya.-B. Zeldovich, Sov Phys. JETP, {\bf 4} 460 (1957).

\ni $^{39}$ F. Cattaneo, ApJ, {\bf 434}, 200 (1994).

\ni $^{40}$ 
J. Stone, E.C. Ostriker, \& C.F. Gammie, ApJ {\bf 508}, L99 (1998).

\ni $^{41}$ W.N. Brandt, T. Boller, A.C. Fabian,\& M. Ruszkowski, 
MNRAS, {\bf 303}, L53 (1999).

\ni $^{42}$ J. Maron, PhD thesis, Caltech, in preparation (2000).

\ni $^{43}$ H. Chou, submitted to ApJ, astro-ph/0009389 (2000).

\ni $^{44}$ D. Balsara \& A. Pouquet, Phys. of Plasmas, {\bf 6} 89 (1999).

\ni $^{45}$ A. Brandenburg, submitted to MNRAS, astro-ph/0011081 (2000)

\ni $^{46}$ E. Vishniac \& J. Cho, submitted to ApJ, astro-ph/0010373 (2000).

\ni $^{47}$ J.F. Hawley, C.F. Gammie, S.A. Balbus, ApJ {\bf 464} 690 (1996).
\end{document}